# Spectrum-free integrated photonic remote molecular identification and sensing

Ross Cheriton[1*], Suresh Sivanandam[2,3], Adam Densmore[4], Ernst De Mooij[5], Daniele Melati[1], Mohsen Kamandar Dezfouli[1], Pavel Cheben[1], Danxia Xu[1], Jens H. Schmid[1], Jean Lapointe[1], Rubin Ma[1], Shurui Wang[1], Luc Simard[4], and Siegfried Janz[1]

[1]*Advanced Electronics and Photonics, National Research Council Canada, Ottawa,, Canada*
[2]*David A. Dunlap Department of Astronomy and Astrophysics, University of Toronto, Toronto, Canada*
[3]*Dunlap Institute for Astronomy and Astrophysics, University of Toronto, Toronto, Canada*
[4]*Herzberg Astronomy and Astrophysics, National Research Council Canada, Victoria, Canada*
[5]*Astrophysics Research Centre, School of Mathematics and Physics, Queen's University Belfast, Belfast, UK*

*Corresponding author: ross.cheriton@nrc-cnrc.gc.ca*

**Abstract:** Absorption spectroscopy is widely used in sensing and astronomy to understand remote molecular compositions. However, dispersive techniques require multichannel detection, reducing detection sensitivity while increasing instrument cost when compared to spectrophotometric methods. We present a novel non-dispersive infrared molecular detection and identification scheme that performs spectral correlation optically using a specially tailored integrated silicon ring resonator. We show experimentally that the correlation amplitude is proportional to the number of overlapping ring resonances and gas lines, and that molecular specificity can be achieved from the phase of the correlation signal. This strategy can enable on-chip detection of extremely faint remote spectral signatures.

## 1. Introduction

Absorption spectroscopy is an important tool for the remote determination of molecular composition of a target gas where direct interaction with a target is not practical or feasible. Typical applications range from the measurement of trace gases in the atmosphere, gas emissions [1], hyperspectral ground [2] and satellite-based [3] remote sensing platforms, to deep-sky and solar system astronomical spectroscopy [4,5]. Many other high sensitivity molecular spectroscopy techniques have been pursued using active techniques such as dual comb spectroscopy [6–11] but these require laser sources, complex electronics and optics, and cannot probe distant targets. In particular, the compositional analysis of astronomical objects and distant atmospheric targets both must rely on simple absorption spectroscopy with naturally occurring ambient light as the only available light source. Nevertheless, many molecular species can be detected and identified by their unique infrared absorption spectrum that results from their ro-vibrational absorption line distributions overlaid upon a broadband background.

Most absorption spectroscopy platforms are built around some form of grating spectrometer that disperses the incoming light across a detector array to capture the spectrum, which is subsequently analyzed to extract the molecular absorption features of interest. In astronomy, modern large telescopes must be matched to correspondingly large grating spectrometers to match their étendues [12]. Replacing full grating spectrometers with compact devices that can detect and quantify the presence of a specific target molecule may significantly reduce the size, cost, mass, development time, and complexity of astronomical instrumentation in applications where a complete spectrum is not required. Such size and mass characteristics are even more considerable advantages for deployment in a space-based telescope. Integrated photonic systems allow for the processing of light in the plane of a centimetre-sized chip. Silicon photonics is among the most developed of these technologies, with many different types of integrated optical devices having been developed for telecommunications and sensing. Examples include integrated filters, modulators, wavelength (de)multiplexers (i.e. spectrometers), optical switches, phase shifters [13], and label-free biosensors [14,15]. Planar waveguide spectrometers/(de)multiplexers can be implemented as echelle gratings [16], arrayed waveguide gratings [17,18], Fourier transform spectrometers [19,19–21], and photonic crystal superprisms [22] on integrated platforms. Waveguide ring resonators have been used as local gas sensors [23,24], but these devices rely on detecting the interaction of trace gases with the evanescent field extending outside the single mode waveguides and cannot be used for remote detection.

The state-of-the-art in optical gas sensing technology has been described in several review papers [11,25,26]. In the case of remote absorption spectroscopy, the detection of absorption features in the spectrum of the input light is used to infer the presence and type of molecular species lying between a broadband light source and the detector. As molecular species exhibit unique spectral fingerprints in the infrared, this uniqueness can be used to identify the molecule through a correlation of the broadband spectrum with a matching spectral filter. We present a novel, non-dispersive method that forgoes spectral acquisition, and instead carries out optical correlation in a fully integrated photonic device through the modulation of a specially designed optical transmission comb. The output of this device is sent to a single channel detector where the correlation signal is encoded as an intensity variation. This principle has been demonstrated using bulk optic Fabry-Pérot (FP) interferometers to produce a periodic transmission filter that can be correlated with a gas spectrum over a finite spectral region [27–31]. The infrared absorption spectrum of many gases exhibit quasi-periodic absorption spectra generated by their coupled vibrational and rotational degrees of freedom. While the absorption line spacings are not perfectly periodic, the periodicity is sufficient over a narrow spectral range to simultaneously overlap numerous transmission lines of a FP interferometer cavity with a suitably chosen cavity length. While the previously mentioned work used FP interferometers [26-29] as the comb filter, waveguide ring resonators can be used for the same purpose since they function as on-chip FP cavities. Silicon waveguide ring resonators are preferable as they are significantly more compact, mechanically stable, have lower power consumption, lower cost, and their spectrum can be tuned at a much higher frequency.

In this paper, we design and demonstrate the first integrated photonic ring resonator remote gas sensor. By thermally modulating a silicon waveguide ring resonator transmission comb spectrum while coupling light from the target through the ring, we obtain the phase and amplitude of the signal for detection and identification, respectively, of the target gas based on the presence of absorption lines in the incoming broadband light signal. This allows for on-chip detection and identification of remote molecular species without spectrum acquisition using only a single detection channel. This sensor analyzes light coupled into the sensor chip through a single mode input optical fibre. The light is coupled from the target source to the input fibre using collimation optics for passive, distant remote gas sensing and identification where active (laser-based) detection is not possible, or by an astronomical telescope coupled to an adaptive optics system. Optical fibres can effectively capture stellar light for further analysis in astronomy, provided the image is approximately diffraction-limited. The output light signal after passing through the sensor can be measured by directing the output light onto a single highly sensitive photodiode that is coupled to the chip either directly using a lens or through an intermediary output fibre. Additional devices can be tailored in a straightforward manner to detect other molecular species (e.g. carbon monoxide, carbon dioxide).

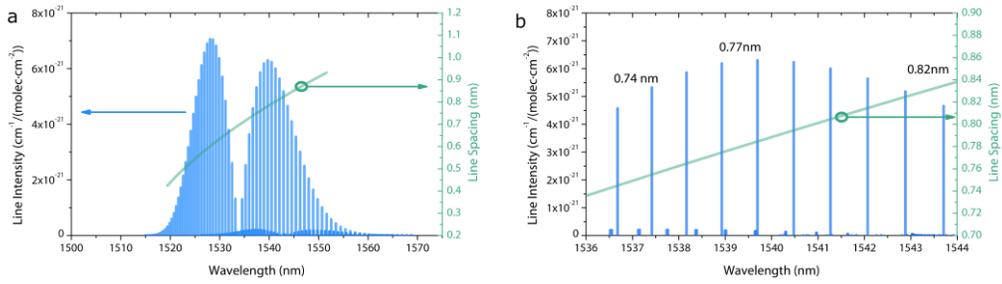

Fig. 1. (a) Ro-vibrational absorption spectrum and line spacings as a function of wavelength for the HCN molecule around 1540 nm (194.67 THz) and (b) the *P*-branch absorption lines and corresponding spectral spacing change with wavelength for the HCN *P*-branch around 1540 nm (194.67 THz).

## 2. Simulation

For this work, we target the most abundant isotope of hydrogen cyanide ($H^{12}C^{14}N$) gas as a proof of concept due to its strong absorption cross-section in the telecommunications C- band, as shown in Fig. 1a. We design a ring resonator as a correlation filter with a free spectral range (FSR) that best matches the line spacing of the strongest HCN absorption lines in the C-band. To achieve the desired FSR of the ring resonator, the ring cavity length and also the targeted spectral lines must be chosen carefully to optimize the difference in overlap of multiple ring resonances and gas lines as the filter is spectrally shifted. The optimization of this matching has two primary benefits: a larger possible correlation signal, and molecule specificity.

We generate the HCN absorption spectrum $A(\lambda)$ using line wavelengths and relative intensities from the high-resolution transmission molecular absorption database (HITRAN) [32] and modeled them as overlapping approximated Voigt functions. An absorption spectrum is created by subtracting the Voigt functions from a flat, normalized background.

The HCN *P*-branch centred at $\lambda = 1540$ nm (194.67 THz) was chosen as the target spectral signature since these lines have a lower relative line spacing change with wavelength than the *R*-branch at shorter wavelengths, and is therefore more suitable for correlating with the more periodic ring resonator transmission spectrum. Note that both the ring spectrum comb and the absorption line spacings are not truly periodic, and also change at different rates with wavelength. The spacing between HCN absorption lines varies from 0.74 nm to 0.82 nm (94 GHz to 105 GHz) for the HCN *P*-branch shown in Fig. 1b. Since the absorption lines at 1540 nm (194.67 THz) are the strongest in the HCN *P*-branch, a target FSR of approximately 0.77 nm (97.5 GHz) is likely to lead to greater detection sensitivity.

Here we outline the simulation of the transmission through a ring resonator filter from its physical and optical characteristics. The resonance vacuum wavelengths $\lambda_m$ and frequencies $\nu_m$ of a ring resonator occur under the condition

$$\lambda_m = \frac{n_e L}{m}, \quad \nu_m = \frac{cm}{n_e L}, \quad m = 1,2,3 \ldots \quad (1)$$

where $n_e$ is the effective index of the waveguide mode, $c$ is the speed of light, $m$ is the longitudinal mode number, and $L$ is the round trip length of the ring. The vacuum wavelength and frequency are related by $\nu = \frac{c}{\lambda}$.

The FSR of a ring resonator in wavelength (frequency) is described by

$$FSR_\lambda = \frac{\lambda^2}{n_g L}, \quad FSR_\nu = \frac{c}{n_g L} \quad (2)$$

where the group index $n_g$ of the waveguide mode. The finesse ($F$) and quality factor ($Q$) are related to the coupling strengths and losses in the resonator, and are defined as:

$$F = \frac{FSR_\nu}{FWHM_\nu}, \quad (3)$$

$$Q = m\frac{FSR_\nu}{FWHM_\nu}. \quad (4)$$

We designed a silicon-on-insulator strip waveguide with a height of 220 nm and width of 450 nm. We calculated the mode properties using a finite difference eigenmode solver (Lumerical Mode Solutions) as a function of wavelength and temperature. We use a temperature and wavelength dependent index of refraction model for silicon of the waveguide to determine effective and group indices [33]. The effective index of the calculated TE mode is $n_e$ ~ 2.37 and a group index of $n_g$ ~ 4.30 at a wavelength of $\lambda$ = 1539 nm (194.8 THz).

The transmission through an add-drop ring resonator is expressed by

$$T_t = \frac{r_2^2 a^2 - 2r_1 r_2 a \cos\theta + r_1^2}{1 - 2r_1 r_2 a \cos\theta + r_1 r_2 a^2} \quad (5)$$

for the through port, and

$$T_d = \frac{(1 - r_1^2)(1 - r_2^2)a}{1 - 2r_1 r_2 a \cos\theta + r_1 r_2 a^2} \quad (6)$$

for the drop port [34].

The parameters $r_1$ and $r_2$ are the coupling coefficients into and out of the ring, respectively, $a$ is the single round trip amplitude transmission for the ring. The temperature dependent phase of the light in the waveguide is described by

$$\theta(T) = \frac{2\pi n_e(T) L}{\lambda}, \quad (7)$$

where $n_e(T)$ is the temperature dependent effective index. Since the TE mode is largely contained inside the silicon waveguide, the spectral shift with temperature can be approximated using the thermo-optic coefficient of silicon. We can use this temperature dependence to dynamically correlate the ring resonator output spectrum to the absorption lines of HCN. The spectral shift is described by

$$\Delta\lambda_{TO} = \sigma_{TO} \lambda_0 \frac{\Delta T}{n_g}, \quad (8)$$

where $\sigma_{TO}$ is the thermo-optic coefficient, $\lambda_0$ is the resonance wavelength and $\Delta T$ is the change in temperature.

The resulting amplitude and relative phase of the integrated intensity can be used to identify and quantify the gas in the incident beam path. The thermo-optic coefficient of silicon is very large at $1.8 \times 10^{-4}$ K$^{-1}$, enabling tuning of the ring resonances with a small change in temperature. When light with an imprinted absorption spectrum is inputted into the ring resonator, the output spectrum of the ring resonator is

$$S_{d,\lambda}(\lambda, T) = A(\lambda) \cdot T_d(\lambda, T), \quad (9)$$

where $A(\lambda)$ is the gas absorption spectrum, and $T_d(\lambda, T)$ is the ring resonator drop port transmission spectrum. The correlation signal $C(T)$ is therefore measured using a single channel detection element with spectrally uniform detection efficiency, and can be expressed by the integral

$$C(T) = \int_{\lambda_1}^{\lambda_2} S_{d,\lambda}(\lambda, T) d\lambda. \quad (10)$$

A suitable ring length can be calculated from Equation 2 by solving for ring length given a target FSR of 0.77 nm (97.5 GHz) that matches the line spacing of a target molecule. From this simple calculation a group index of 4.34 leads to a ring length of ~709 µm. We verify the performance of a ring resonator with this ring length ($L$) by calculating the maximum correlation signal amplitude for ring lengths ranging from 400 µm to 1200 µm. Ring resonator drop port transmission spectra as a function of temperature and wavelength are correlated with an HCN absorption spectrum using Equation 9. The spectral products at $S_d$ are then integrated using Equation 10 to produce the correlation signal, as shown in Fig. 2a. We use the ring resonators parameters shown in Table 1. We choose *a=0.9* as it is a typical silicon-on-insulator single round trip amplitude transmission coefficient in fabricated ring resonators. The values for $r_1$ and $r_2$ are chosen to approximate a moderate coupling strength to the ring resonator.

**Table 1. Simulation Parameters**

| Parameter | Value | |
|---|---|---|
| $a$ | 0.9 | |
| $r_1$ | 0.9 | |
| $r_2$ | 0.9 | |
| Quality factor | 19,000 | |
| $n_g$ | 4.30 | |
| $n_e$ | 2.37 | |
| Finesse | 8.5 | |
| Ideal ring round trip length | 705 µm | |
| FWHM of gas lines | 30 pm | (38 GHz) |
| Bandpass filter | 0.75-6 nm | (95 GHz-760 GHz) |

The bounds of the integral in Equation 10 are finite since not the ring resonances cannot simultaneously overlap with all the absorption lines of HCN. The bandpass is practically limited to roughly 6 nm (760 GHz), but is reduced with high quality factors (sharper ring resonances). Beyond a 6 nm bandpass, additional absorption lines cannot overlap simultaneously and will only serve to broaden, and eventually reduce, the amplitude of the correlation function with a spectral shift of the ring transmission spectrum [35]. Therefore, we impose an external bandpass filter in our simulations. These calculations were repeated for bandpass widths ($\Delta\lambda_b = \lambda_2 - \lambda_1$) from 0.75 nm to 6 nm (95 GHz to 760 GHz). At a ring round trip length of 705 µm, with bandpass filter widths of 1.5 nm (190 GHz) and larger, the modulation amplitude reaches a maximum similar to our original estimate of 709 µm. As the bandpass filter width increases, the signal contrast between aligned and misaligned conditions (correlation amplitude) around 705 µm increases. This arises from the simultaneous overlap of multiple ring resonances with multiple gas absorption lines and a corresponding higher sensor sensitivity. Other ring lengths also display weaker correlation features in modulation amplitude, such as at 470 µm and 1060 µm, when the ring resonances overlap a subset of the ring resonances in the bandpass.

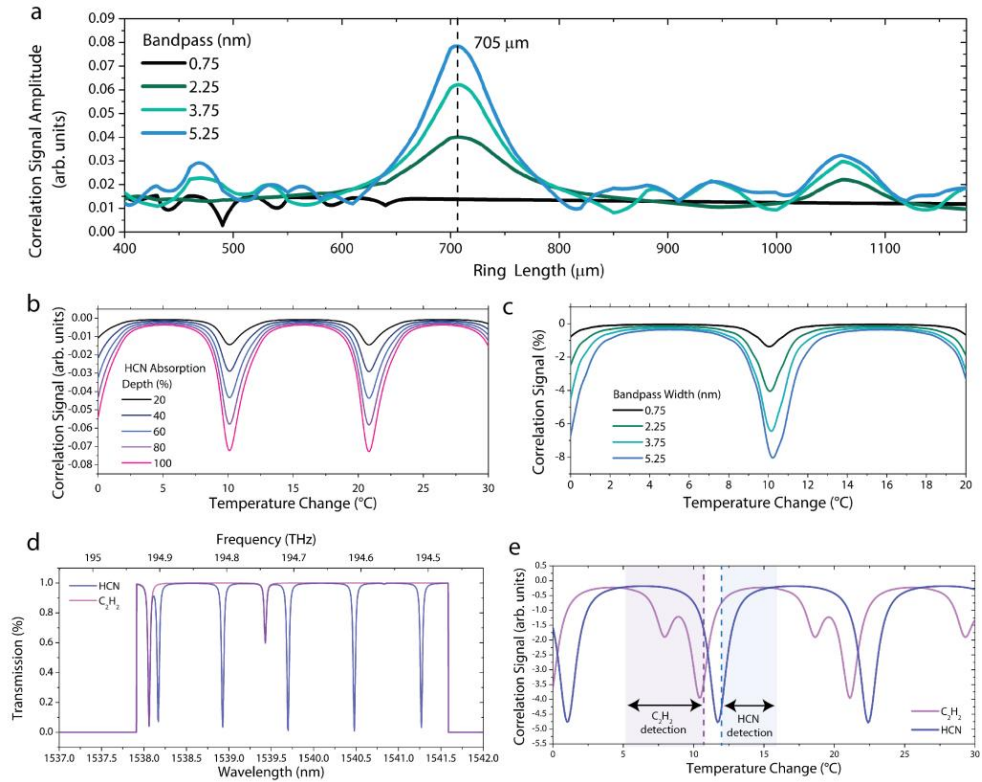

Fig. 2. (a) Simulated correlation amplitude as a function of ring round trip length and bandpass filter showing an optimal ring length of 705 μm for HCN detection. (b) Simulated correlation signal as a function of HCN absorption line depth for a bandpass of 6 nm (760 GHz). (c) Simulated correlation signal as a function of bandpass filter width. (d) Normalized $C_2H_2$ and HCN absorption spectrum over a 3.75 nm (450 GHz) bandpass. (e) Simulated correlation signal for $C_2H_2$ and HCN over a 3.75 nm (450 GHz) bandpass. The temperature offsets and spectral modulation regions that can be used obtaining specificity between $C_2H_2$ and HCN are highlighted.

With HCN absorption features present in the input spectrum, simulated correlation signal minima are observed every 11°C in temperature change, or a shift of approximately 0.77 nm (97.5 GHz) of the ring transmission spectrum. In Fig. 2b, we show the correlation signal as a function of temperature change for various HCN absorption depths with a bandpass of 6 nm (760 GHz). The values of the minima in the correlation signal decrease since deeper absorption lines from HCN increasingly block more light at the drop port of the ring, showing that the sensor can distinguish relative HCN column density in the beam path. We also investigate the effect of the external bandpass filter width, which allows for more gas lines to contribute the signal. The correlation signal is plotted as a function of temperature at different bandpass widths in Fig. 2c, showing increasing signal contrast with larger bandpasses up to 6 nm (760 GHz). This is due to the simultaneous overlapping of additional absorption lines with ring resonances, resulting in an increase in the absolute correlation signal amplitude. The signal is also broadened due to imperfect overlap of the gas lines with the ring resonances. As the gas lines are asymmetrically distributed, a slightly asymmetrical shape to the correlation signal is visible at the large bandpass filter widths.

We also investigate the molecular specificity of our sensor, since multiple gases can have absorption features in the bandpass. In the case of acetylene ($C_2H_2$) present with HCN, we now have two additional non-periodic strong absorption lines which complicate our spectrum, as shown in Fig. 2d. The correlation of the $C_2H_2$ spectrum with the ring drop port transmission reveals a new correlation signal pattern. Fig. 2e shows that correlation features occur at different temperatures, one corresponding to $C_2H_2$, and the other to HCN. As in this case, it is possible to judiciously choose a filter centre wavelength to ensure non-overlapping spectral features from different molecular species. The resulting correlation features that occur at unique ring temperatures, providing greater molecular specificity without acquiring a spectrum.

The arrows in Fig. 2e indicate possible temperature modulation offsets such that the same sensor can independently detect either $C_2H_2$ or HCN. We expect this molecular specificity feature to be upheld for other gases if the strongest absorption lines of both gases do not overlap. Otherwise, molecular specificity can also be achieved by observing the change in the correlation signal with increasing bandpass, since the correlation signal from $C_2H_2$ arises from only two absorption lines and a mismatched ring FSR.

The performance of the sensor depends on the ring resonator coupling coefficients $\underline{a}$, $r_1$ and $r_2$. Higher values of the single-pass amplitude transmission ($a>0.95$) correspond to less loss (e.g. less waveguide sidewall roughness) in the fabricated ring resonator, and generally leads to more light coupled into the drop port. However, obtaining high spectral contrast of the ring resonator filter relies on being near the critical coupling condition which total ring round trip loss including drop port coupling is equal to the through port coupling. The $r_1$ and $r_2$ parameters are set based on the coupling section lengths and coupling gaps (higher $r_1$ and $r_2$ values enable higher quality factors) and produce critical coupling conditions if $r_1=r_2a$. We have plotted the correlation signal as a function of temperature for a range of quality factors under critical coupling conditions in Fig. 3a. The correlation amplitude will depend on the gas absorption line and ring resonance overlap across the sampled spectral passband. In an ideal correlation filter, the gas spectrum and filter spectrum should match exactly, but this is not possible using a simple ring filter. Since the gas line spacings and ring transmission line spacings are not exactly commensurate, the optimal choice of ring resonance linewidth therefore depends on the difference in gas and ring line spacings, the spectral linewidths, and the sampled spectral bandwidth.

The peak-to-peak modulation of the correlation signal is maximized at $r_1=0.67$ with $r_2=0.7$ as shown in Fig. 3b for a 6 nm bandpass window. This corresponds to a quality factor for the sensor of ~9000 with $a=0.95$. With even lower quality factors, the resonance peak-to-minimum transmission contrast is reduced. At higher quality factors, the ring resonances are sharper and overlap with gas absorption lines is less leading to reduced peak-to-peak correlation signal amplitudes. For high pressure, high temperature target gases, even lower quality factors are better suited to accommodate the broadened spectral features.

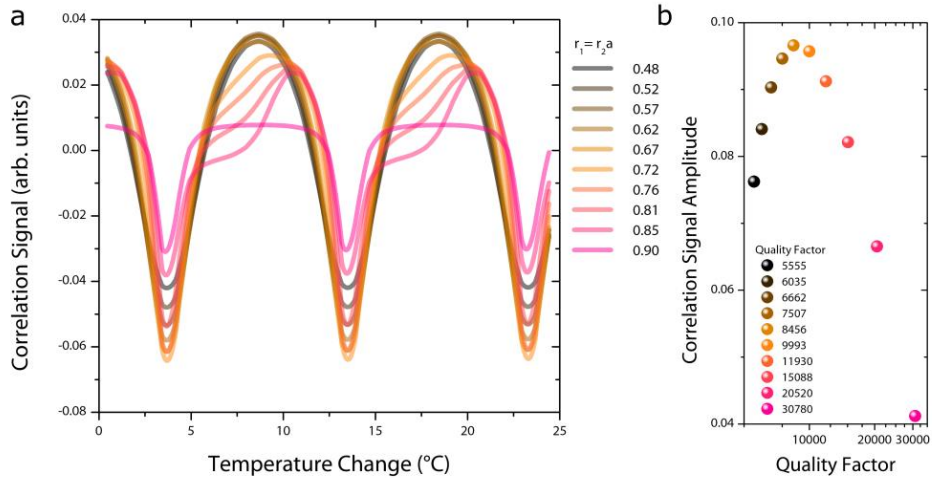

Fig. 3. (a) Simulated correlation signal as a function of temperature change for a bandpass spectral width of 6 nm (760 GHz). The single pass amplitude transmission parameter is fixed at $a = 0.95$ and coupling coefficients $r_1$ and $r_2$ are varied while maintaining the critical coupling condition ($r_1 = r_2a$). (b) Simulated correlation signal amplitude (peak-to-peak) for ring quality factors corresponding to the ring parameters given in Fig 3a.

### 3. Fabrication and experimental setup

The silicon ring resonator devices are fabricated on the silicon-on-insulator (SOI) platform using a JEOL JBX-8100FS E-beam lithography tool through Applied Nanotools. The ring resonator consists of a silicon waveguide formed into a racetrack loop. The entire chip with input and output coupling fibres is shown in Fig. 4a, with the ring resonator coupling section shown in detail in Fig. 4b, respectively. The buried oxide layer is 2 µm thick with a 220 nm thick silicon layer. Devices are fully etched to the buried oxide and have waveguide dimensions of 450 nm wide and 220 nm high. The waveguides are clad with a 2.2 µm thick oxide layer. TiW microheaters are patterned atop the $SiO_2$ cladding to provide local heating to the waveguides. All heaters are covered in 300 nm of $SiO_2$ and provide a tuning rate of approximately 0.6 °C/mA at low currents with a heater resistance of ~100 Ω. The heaters are kept short due to the high resistivity of TiW and to support sufficient heating at low voltages. Light input and output are achieved through subwavelength edge couplers designed for adiabatic, low-loss broadband light coupling from a tapered fibre to the waveguide around 1550 nm (~193 THz) [36,37]. A deep trench etch at the chip edges through the entire wafer thickness is used create provide smooth input and output waveguide facets. Devices are designed for TE polarization to increase the thermo-optic coefficient of the waveguides which is proportional to the mode overlap ratio between silicon and $SiO_2$. Coupling between the input waveguide and the racetrack ring resonator is accomplished using directional couplers with a closest coupling gap of 200 nm. A tapered waveguide is used in the drop output waveguide to reduce back-reflections from counter-propagating modes. The direction couplers were optimized with 5.4 µm and 6 µm parallel sections for the drop and through coupling sections, respectively. The asymmetric coupling length is designed to accommodate for some expected loss in the add-drop ring to better approach the critical coupling condition [30]. The coupling lengths were designed to produce a ring resonator with a low quality factor while still providing a reasonable transmission contrast between off- and on-resonance light. Other devices with symmetric coupling configurations produced resonances with less resonance transmission contrast. The racetrack ring bend radii are fixed at 20 µm to ensure bend losses in the ring are negligible. The propagation loss of the TE mode in the waveguides was ~1.3 dB/cm, which leads to an estimated intrinsic ring resonator quality factor of 680,000. Different rings lengths were patterned to accommodate for variations in fabrication. A ring length a 737 µm led to the best overlap with the HCN lines. This length differs slightly from the optimal length of 709 µm found in the previous section, presumably due to differences between the fabricated and the designed waveguide geometry and material constants.

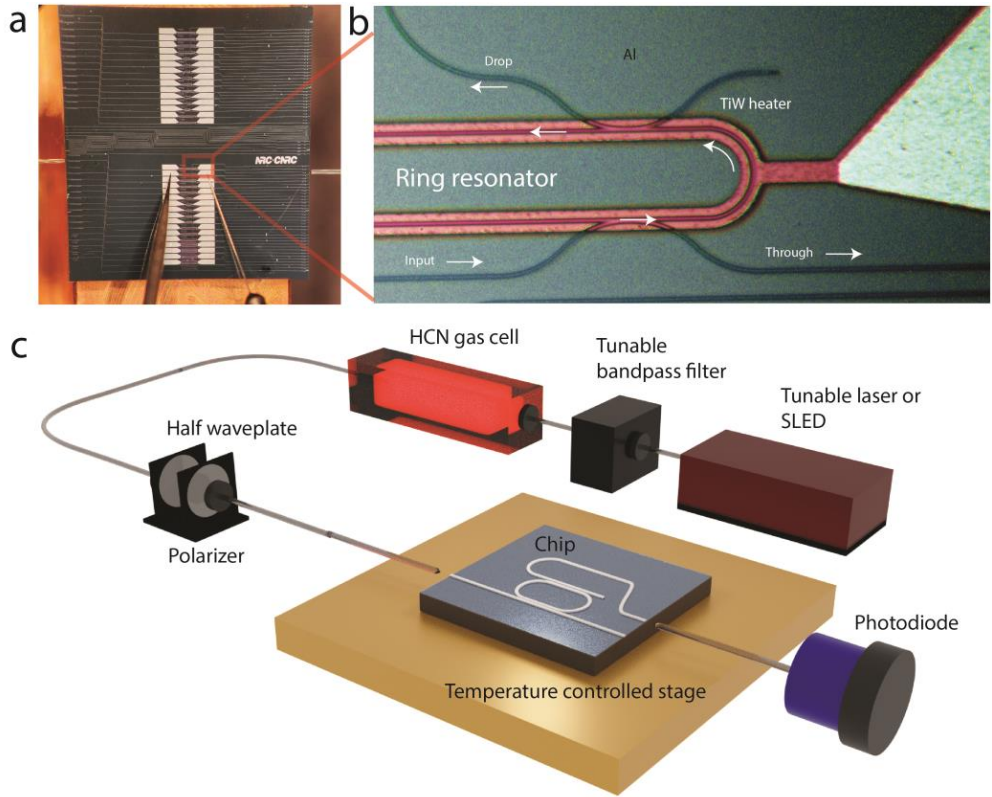

Fig. 4. (a) Plan-view optical photograph of the fabricated photonic chip with tapered fibre input and output couplers and two electrical probes contacting the electrical pads. (b) Optical micrograph of a fabricated racetrack ring resonator (dark), microheater (red), and contact pads (light green area to the right). (c) Illustration of the experimental optical characterization setup.

The chips are placed on a temperature controlled copper stage as shown in Figs. 4a and 4c. The temperature of the stage is held constant while varying the current through the microheaters to modulate the drop port output spectrum. Polarization maintaining fibres are used for light input and output coupling to ensure the light entering the ring resonators are TE polarized. The contact pads of the microheater are contacted with needle probes connected to a current source. The coupling section of the racetrack ring resonator is shown with microheaters after fabrication in Fig. 4b. A schematic of the optical characterization system is shown in Fig. 4c. The transmission spectra of the ring and gas cell are acquired using a tunable laser source and photodiode. The light polarization is controlled using a polarization controller that is adjusted to couple only TE polarized light to the chip. An $H^{12}C^{14}N$ gas cell at a pressure of 10 Torr with a path length of 5.5 cm is used to produce absorption line features.

The optical set-up is also alternatively configured with a broadband light source, a superluminescent diode system centred at 1555 nm (192.79 THz). In this case, the input light is passed through a tunable bandpass filter, enabling tuning of the centre wavelength and spectral width of the bandpass. In this configuration, the photodiode measures the total light power exiting the drop port. This configuration involves incoherent light, and enables the direct measurement the correlation signal, and therefore better represents more typical operating characteristics and conditions of the sensor.

4. **Results**

*A. Detection using a tunable laser*

The racetrack ring resonators were first characterized from 1534 nm to 1543 nm (195.43 to 194.29 THz) at the through and drop ports using the tunable laser. The transmission characteristics of the drop outputs for a 737 µm long ring are shown in Fig. 5a. The FWHM of the drop port resonances are found to be ~110 pm (~14 GHz) from Fig. 5b. Resonance depths are about -13 dB, corresponding to a quality factor of about 14,000 with a finesse of ~6.8 for the ring resonator. The simulated ring finesse in the previous section was ~9.5, which is a slightly less ideal ring coupling condition when compared to the fabricated ring resonator. The HCN gas cell produces very sharp absorption line features with linewidths of ~8 pm (~1 GHz) at full-width at half minimum (FWHM). Broader linewidths for the gas lines would have contributed to a stronger correlation signal amplitude.

While the correlation technique can be applied either using the through or drop ports (or potentially both simultaneously), the drop port signal provides a higher signal to background ratio since the integrated power is less susceptible to interference and other spurious effects across the chip and measurement system. However, in a more complex two output configuration, both drop and through port signals can be used to provide intensity normalization. The gas cell is removed from the fibre-optic system when obtaining the reference correlation signals shown in Fig. 5c. With the HCN gas cell inserted into the light path, absorption features are overlaid upon the drop port transmission spectrum of the ring resonator. For these experiments, the stage temperature is maintained at 20 °C. The drop port transmission spectrum is red-shifted by increasing the current through the microheater as shown in Fig. 5a and 5b. The resonances shift by the full free spectral range of the ring approximately every 11°C. The temperature dependent transmission spectrum of a single drop port resonance mode as it crosses a single HCN absorption line at 1539.7 nm (194.7084 THz) is shown in Fig. 5b. At a microheater current of around 15-18 mA, the ring is heated to ~6 °C above the stage temperature and the resonances are aligned with the HCN absorption lines. The narrow width of the HCN line leaves a sharp drop in transmitted power at specific wavelengths.

In order to emulate the sensor operation using background broadband illumination, we integrate (using Equation 10) the tunable laser transmission spectra through the ring drop port across the desired passband where the gas lines are strongest and best match the drop port resonances. The measurements were repeated when the gas cell was removed, providing a reference spectrum representing the reference zero-gas correlation as a function of microheater current. The correlation signals without the gas cell (ring only), shown in Fig. 5c, are normalized for direct comparison at bandpass values from 0.75 nm to 6 nm (95 GHz to 760 GHz), spanning 1 to 8 HCN absorption lines.

With increasing bandpass widths, the correlation feature at 18 mA is more apparent and becomes broader. As is evident from Fig. 5d, the contributions from multiple ring resonances overlapping with multiple gas lines leads to an improvement in the detection. The broadening of the correlation signal minimum is a result of the mismatch in aperiodicity of the gas lines and the ring resonances.

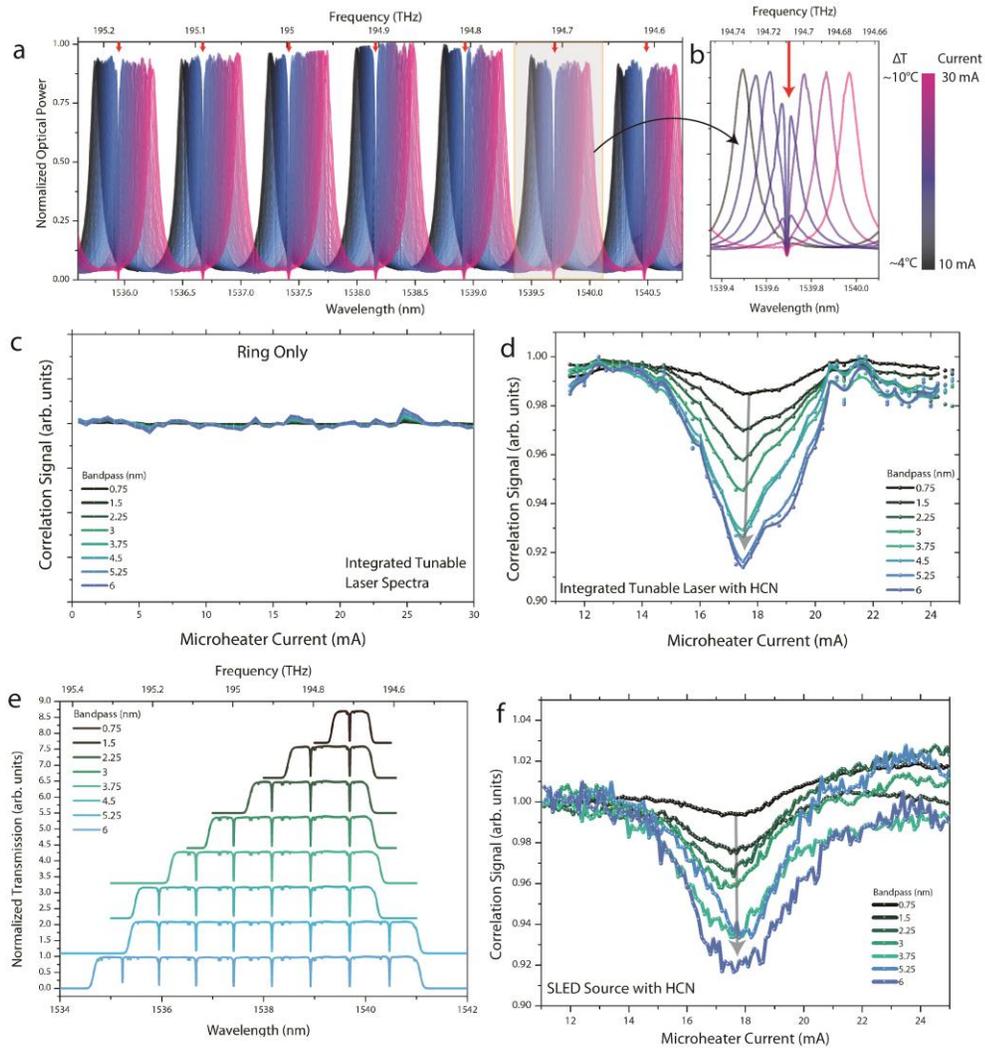

Fig. 5. (a) Experimental ring resonator drop port transmission spectrum with HCN gas absorption features as measured by a scanning tunable laser source as a function of ring temperature. The change in colour indicates the change in ring temperature. (b) High resolution ring resonator drop port transmission spectrum of a single resonance and gas absorption line. (c) Integrated tunable laser spectra through the ring drop port without an HCN gas cell, and (d) with an HCN gas cell with varying bandpass filter widths. (e) Transmission spectra through the HCN gas cell at different bandpass filter widths. (f) Difference between correlation signals with and without the HCN gas cell at the ring resonator drop port as a function of microheater current using the broadband LED source.

## B. Demonstration using broadband light

We then demonstrate sensing using a weak broadband background provided by a superluminescent LED. The spectral power density detected at the photodiode is around 30 pW/nm (0.2 pW/GHz), or approximately >200,000 weaker than the tunable laser. This source better represents a dim target typically encountered in astronomical observations. Different bandpass widths and centre wavelengths are selected using a tunable external bandpass filter to accommodate 1 to 8 gas lines. The normalized transmission spectra through the filter and the gas cell for varying bandpass widths are measured using a tunable laser sweep and are shown in Fig. 5e. The microheater current is increased from 11 mA to 25 mA as before, allowing the ring resonances to simultaneously align (and misalign) with the gas lines. The power at the photodiode, which inherently integrates the output spectrum, is measured as a function of microheater current with a fixed stage temperature. The difference between the correlation signal with and without the gas cell in the light path is shown in Fig. 5f. When the HCN gas cell is connected, a clear dip by about 8% in the correlation signal is seen centred at ~17-18 mA of microheater current, similar to what is seen by integrating the tunable laser spectra, representing the presence of HCN. The signal-to-noise ratio is estimated at ~250, which is significantly lower due to the weak spectral power density of the SLED source as well as from additional insertion losses from the tunable bandpass filter. When the ring resonances are misaligned with respect to the absorption lines of HCN (below 14 mA and above 22 mA), the output power is at the nominal value corresponding to no gas detection. The correlation signal is similarly broadened and deeper as the bandpass filter width is increased. The overlap of multiple absorption lines is inferred from the deepening of the minimum at 17-18 mA with increasing bandpass filter widths. At wider bandpass filter widths, simultaneous overlap of all ring resonances and gas lines in the bandpass is not possible, and partial overlaps occurs. This effect broadens the signal and results in a limit to the sensitivity when using nearly equally spaced ring resonances.

## 5. Discussion

We describe an integrated photonic remote gas sensor using correlation-based detection and identification of absorption features of a gas contained within a broadband background light. By engineering a silicon waveguide ring resonator with appropriate length and group index, a integrated correlation filter is matched to gases with quasi-periodic absorption features over a bandpass of a few nanometres. We show that the overlap of the ring resonator drop port with the absorption lines of HCN produces a unique modulation pattern that identifies HCN based on the phase of the modulated correlation signal. We additionally demonstrate HCN detection with a weak broadband source. While demonstrated with HCN, many other gases with similar periodic absorption features such as $CO_2$ and CO can be detected similarly with different ring lengths. We also show by simulation that the sensor can operate in the presence of $C_2H_2$ and should distinguish other gases with non-overlapping absorption features in the same bandpass. The sensor performs best with a bandpass filter of 6 nm (760 GHz) where up to 8 resonances contribute to the sensor response, with smaller bandpass filters leading to lower response. Simulations with larger bandpasses reveal negligible gain in signal amplitude with corresponding loss of contrast. In the case of saturated absorption lines, the sensor can still detect the presence of the gas if there is still some light in the bandpass, however, the saturation will decrease our ability to recover the relative column depth for saturated lines.

Integrated correlation gas sensing has multiple advantages over dispersive spectroscopy regarding sensitivity for astronomy and remote sensing. Firstly, in a dispersive spectrometer, the light is divided by wavelength onto multiple pixels, reducing the photon flux per pixel. Whereas in our integrated correlation sensor, we collect all photons into a single detection element (or pixel). In the shot noise dominated condition, the signal-to-noise ratio is related to the photon count ($N_{ph}$) by

$$SNR = \sqrt{N_{ph}}.$$

To resolve the HCN absorption lines over 6 nm of spectral bandwidth for HCN, a dispersive spectrometer would realistically require at least ~30 pixels, reducing the SNR by a factor of ~5 when compared to the integrated correlation ring resonator. Considering the modulation of the ring spectrum across the absorption lines reduces the signal by a factor of ~2, the SNR at the photodiode is still a factor of ~3.5 higher than that of dispersive spectrometers.

Secondly, the use of a single channel allows for the use of larger, higher performance detectors which can be optimized for higher quantum efficiencies, lower cost, lower dark currents, and no pixel cross-talk.

Finally, the rapid modulation of the microheater current allows for lock-in amplification of the resulting signal that is not practically achieved for a highly multichannel detector arrays. A silicon waveguide ring filter transmission can be modulated at several kHz using thermal tuning as in this work, or at up to MHz frequencies and higher rates using carrier-based modulation [38]. The phase sensitive lock-in detection eliminates any source of noise outside of a narrow measurement bandwidth centred at the modulation frequency, including atmospheric fluctuations. Using lock-in techniques, signals many orders of magnitude smaller than the noise background can be routinely acquired, unlike in dispersive spectroscopy. Long time constants for the lock-in detection can be used for signal extraction with extremely low signal-to-noise ratios since detector response time is not usually a priority for astronomical observations. The integrated aspect of this correlation sensor allows for low-cost, compact and mechanical stability not found in bulk optic FP-based implementations [29].

The minimum sensitivity for the ring resonator correlation sensor can be related to the minimum HCN absorption depth detectable with lock-in amplification. We calculate this in the same manner as Vargas-Rodríguez et al. [29] where

$$SNR = 1 = \frac{AM_{RMS}^V}{(NEV + N_{lock})\sqrt{\Delta f}}.$$

We can calculate the minimum lock-in detection signal for a Zurich Instruments MFLI lock-in amplifier with an input noise amplitude of $N_{lock}$ = 40 nV/$\sqrt{Hz}$ at 1 Hz, and assume a reasonable noise equivalent voltage (NEV) of 100 nV/$\sqrt{Hz}$ at 1 Hz. As in [29], we assume the noise equivalent bandwidth is $\Delta f \approx 1/(4\tau_c)$, where the $\tau_c$ is the time constant of the lock-in amplifier. For extremely faint object astronomy, time constants as long as several minutes can be used. Using a lock-in detection simulation for varying optical depths of HCN, we achieve a detection efficiency of ~5 x $10^{-4}$ V per percent optical depth. With $\tau_c$=10 seconds and incorporating the input noise, we have a minimum detectable absorption depth on the order of $10^{-6}$ %. This can be lowered through the use of longer time constants and the use of ultra-low noise detectors. As our proposed optical detection strategy relies on a distant, natural continuum source with intervening gas, we define our lower detection limit in regard to the absorption depth rather than the gas concentration in the column.

In practice, the single TE-optimized ring resonator remote photonic sensor will only be able to operate using half of the incoming incoherent light. Therefore an additional ring designed for TM polarization in a similar fashion should be used in parallel. A polarizing beam splitter implanted on-chip can divide incoming light into both ring resonators appropriately. This also enables polarization sensitivity of the incoming light if the output channels are separate. Another consideration involves the coupling light into a single mode fibre. Efficiently coupling light into a single mode fibre generally requires adaptive optics [39] or space-based diffraction-limited operation. When diffraction-limited performance is not available, one can use a photonic lantern to convert light from a multi-mode fibre that more efficiently couples light into multiple parallel ring resonators.

The primary disadvantage to this technique is the lack of spectral information. It is possible that false positives can be introduced with absorption line combinations from different gases. With higher specificity sensors (e.g. multiple rings or custom Bragg filters), signal contributions from such possible combinations can be suppressed. Simulations under a wide array of absorption line contributions can be used to anticipate these effects.

Another possibility for improved sensitivity and specificity is through dispersion engineering of the ring resonator such that the overlap between resonances and absorption lines is possible over a larger bandpass. Unfortunately, it is difficult to achieve the required extremely high group velocity dispersions to match most gas absorption line distributions. Such dispersion could potentially be engineered through photonic crystal structures operating near the photonic band edge, but in practice this leads to prohibitively high propagation losses.

In this work, ring resonators have been used as convenient device to generate a quasi-periodic comb filter. It also possible to create much more complex waveguide filters based on gratings to precisely match more complex gas spectra over wider spectral ranges, for example using layer peeling methods for Bragg gratings on the silicon nitride platform [40,41]. The precise matching of filter transmission to gas spectra should further increase the specificity and signal to noise of the correlation method at the cost of chip area and more power consumption for thermal modulation.

Our correlation technique reduces the detection requirement from a 1D photodetector array of typical dispersive spectrometer to a single channel, which introduces cost and sensitivity advantages while maintaining molecular specificity as used in full spectrum acquisitions. In extremely low signal applications, avalanche photodiodes or photomultiplier tubes can be used. The multiplexed advantage of using a single channel for multi-gas sensing can also pave the way towards a form of 2D gas mapping where arrays of ring resonators can process light from individual pixels as a form of hyperspectral imaging. Possible applications range from low-cost remote monitoring of gas emissions to enabling astronomical observations of extremely weak absorption features such as exoplanet atmospheres during transits. This strategy can be employed on many other gases which exhibit unique correlation signals with a suitable integrated resonant cavity. Future work includes investigating devices which operate in the mid-infrared where device sensitivity can be significantly higher due to stronger absorption features.

## Disclosures

The authors declare no conflicts of interest.

## References

1. B. Galle, J. Samuelsson, B. H. Svensson, and G. Börjesson, "Measurements of Methane Emissions from Landfills Using a Time Correlation Tracer Method Based on FTIR Absorption Spectroscopy," Environ. Sci. Technol. **35**, 21–25 (2001).
2. H.-J. Park, J.-S. Park, S.-W. Kim, H. Chong, H. Lee, H. Kim, J.-Y. Ahn, D.-G. Kim, J. Kim, and S. S. Park, "Retrieval of NO2 Column Amounts from Ground-Based Hyperspectral Imaging Sensor Measurements," Remote Sensing **11**, 24 (2019).
3. Z. Lee and K. L. Carder, "Absorption spectrum of phytoplankton pigments derived from hyperspectral remote-sensing reflectance," Remote Sensing of Environment **89**, 361–368 (2004).
4. M. A. Cordiner, S. N. Milam, N. Biver, D. Bockelée-Morvan, N. X. Roth, E. A. Bergin, E. Jehin, A. J. Remijan, S. B. Charnley, M. J. Mumma, J. Boissier, J. Crovisier, L. Paganini, Y.-J. Kuan, and D. C. Lis, "Unusually high CO abundance of the first active interstellar comet," Nature Astronomy (2020).
5. Y. J. Pendleton, S. A. Sandford, L. J. Allamandola, A. G. G. M. Tielens, and K. Sellgren, "Near-Infrared Absorption Spectroscopy of Interstellar Hydrocarbon Grains," The Astrophysical Journal **437**, 683 (1994).
6. I. Coddington, N. Newbury, and W. Swann, "Dual-comb spectroscopy," Optica **3**, 414–426 (2016).
7. I. Coddington, W. C. Swann, and N. R. Newbury, "Coherent dual-comb spectroscopy at high signal-to-noise ratio," Physical Review A **82**, 043817 (2010).
8. T. Ideguchi, A. Poisson, G. Guelachvili, N. Picqué, and T. W. Hänsch, "Adaptive real-time dual-comb spectroscopy," Nature communications **5**, 3375 (2014).
9. M. Yu, Y. Okawachi, A. G. Griffith, N. Picqué, M. Lipson, and A. L. Gaeta, "Silicon-chip-based mid-infrared dual-comb spectroscopy," Nature communications **9**, 1–6 (2018).
10. N. Picqué and T. W. Hänsch, "Frequency comb spectroscopy," Nature Photonics **13**, 146–157 (2019).
11. K. C. Cossel, E. M. Waxman, I. A. Finneran, G. A. Blake, J. Ye, and N. R. Newbury, "Gas-phase broadband spectroscopy using active sources: progress, status, and applications [Invited]," J. Opt. Soc. Am. B **34**, 104–129 (2017).
12. D. S. Rupke, S. Veilleux, and D. Sanders, "Keck absorption-line spectroscopy of galactic winds in ultraluminous infrared galaxies," The Astrophysical Journal **570**, 588 (2002).
13. M. Pu, L. Liu, W. Xue, Y. Ding, L. H. Frandsen, H. Ou, K. Yvind, and J. M. Hvam, "Tunable microwave phase shifter based on silicon-on-insulator microring resonator," IEEE Photonics Technology Letters **22**, 869–871 (2010).
14. D.-X. Xu, M. Vachon, A. Densmore, R. Ma, A. Delâge, S. Janz, J. Lapointe, Y. Li, G. Lopinski, and D. Zhang, "Label-free biosensor array based on silicon-on-insulator ring resonators addressed using a WDM approach," Optics letters **35**, 2771–2773 (2010).
15. T. Claes, J. G. Molera, K. De Vos, E. Schacht, R. Baets, and P. Bienstman, "Label-free biosensing with a slot-waveguide-based ring resonator in silicon on insulator," IEEE Photonics journal **1**, 197–204 (2009).

16. S. Janz, A. Balakrishnan, S. Charbonneau, P. Cheben, M. Cloutier, A. Delâge, K. Dossou, L. Erickson, M. Gao, and P. A. Krug, "Planar waveguide echelle gratings in silica-on-silicon," IEEE Photonics technology letters **16**, 503–505 (2004).
17. P. Cheben, J. H. Schmid, A. Delâge, A. Densmore, S. Janz, B. Lamontagne, J. Lapointe, E. Post, P. Waldron, and D.-X. Xu, "A high-resolution silicon-on-insulator arrayed waveguide grating microspectrometer with sub-micrometer aperture waveguides," Optics express **15**, 2299–2306 (2007).
18. H. Takahashi, S. Suzuki, K. Kato, and I. Nishi, "Arrayed-waveguide grating for wavelength division multi/demultiplexer with nanometre resolution," Electronics letters **26**, 87–88 (1990).
19. D. M. Kita, B. Miranda, D. Favela, D. Bono, J. Michon, H. Lin, T. Gu, and J. Hu, "High-performance and scalable on-chip digital Fourier transform spectroscopy," Nature communications **9**, 1–7 (2018).
20. K. Okamoto, H. Aoyagi, and K. Takada, "Fabrication of Fourier-transform, integrated-optic spatial heterodyne spectrometer on silica-based planar waveguide," Optics letters **35**, 2103–2105 (2010).
21. H. Podmore, A. Scott, P. Cheben, A. V. Velasco, J. H. Schmid, M. Vachon, and R. Lee, "Demonstration of a compressive-sensing Fourier-transform on-chip spectrometer," Optics letters **42**, 1440–1443 (2017).
22. H. Kosaka, T. Kawashima, A. Tomita, M. Notomi, T. Tamamura, T. Sato, and S. Kawakami, "Superprism phenomena in photonic crystals," Physical review B **58**, R10096 (1998).
23. T. H. Stievater, M. W. Pruessner, D. Park, W. S. Rabinovich, R. A. McGill, D. A. Kozak, R. Furstenberg, S. A. Holmstrom, and J. B. Khurgin, "Trace gas absorption spectroscopy using functionalized microring resonators," Opt. Lett. **39**, 969–972 (2014).
24. G. Mi, C. Horvath, and V. Van, "Silicon photonic dual-gas sensor for H2 and CO2 detection," Opt. Express **25**, 16250–16259 (2017).
25. A. Hänsel and M. J. R. Heck, "Opportunities for photonic integrated circuits in optical gas sensors," Journal of Physics: Photonics **2**, 012002 (2020).
26. J. Hodgkinson and R. P. Tatam, "Optical gas sensing: a review," Measurement Science and Technology **24**, 012004 (2012).
27. J. J. Barrett and S. A. Myers, "New Interferometric Method for Studying Periodic Spectra Using a Fabry–Perot Interferometer*," J. Opt. Soc. Am. **61**, 1246–1251 (1971).
28. J. Kuhn, U. Platt, N. Bobrowski, and T. Wagner, "Towards imaging of atmospheric trace gases using Fabry–Pérot interferometer correlation spectroscopy in the UV and visible spectral range," Atmos. Meas. Tech. **12**, 735–747 (2019).
29. E. Vargas-Rodríguez and H. N. Rutt, "Design of CO, CO2 and CH4 gas sensors based on correlation spectroscopy using a Fabry–Perot interferometer," Sensors and Actuators B: Chemical **137**, 410–419 (2009).
30. J. Kuhn, N. Bobrowski, P. Lübcke, L. Vogel, and U. Platt, "A Fabry–Perot interferometer-based camera for two-dimensional mapping of SO2 distributions," Atmos. Meas. Tech. **7**, 3705–3715 (2014).
31. Y. Qu, Z.-H. Kang, Y. Jiang, and J.-Y. Gao, "Multiline absorption spectroscopy for methane gas detection," Appl. Opt. **45**, 8537–8540 (2006).
32. I. E. Gordon, L. S. Rothman, C. Hill, R. V. Kochanov, Y. Tan, P. F. Bernath, M. Birk, V. Boudon, A. Campargue, K. V. Chance, B. J. Drouin, J.-M. Flaud, R. R. Gamache, J. T. Hodges, D. Jacquemart, V. I. Perevalov, A. Perrin, K. P. Shine, M.-A. H. Smith, J. Tennyson, G. C. Toon, H. Tran, V. G. Tyuterev, A. Barbe, A. G. Császár, V. M. Devi, T. Furtenbacher, J. J. Harrison, J.-M. Hartmann, A. Jolly, T. J. Johnson, T. Karman, I. Kleiner, A. A. Kyuberis, J. Loos, O. M. Lyulin, S. T. Massie, S. N. Mikhailenko, N. Moazzen-Ahmadi, H. S. P. Müller, O. V. Naumenko, A. V. Nikitin, O. L. Polyansky, M. Rey, M. Rotger, S. W. Sharpe, K. Sung, E. Starikova, S. A. Tashkun, J. V. Auwera, G. Wagner, J. Wilzewski, P. Wcisło, S. Yu, and E. J. Zak, "The HITRAN2016 molecular spectroscopic database," Journal of Quantitative Spectroscopy and Radiative Transfer **203**, 3–69 (2017).
33. D.-X. Xu, A. Delâge, P. Verly, S. Janz, S. Wang, M. Vachon, P. Ma, J. Lapointe, D. Melati, P. Cheben, and J. H. Schmid, "Empirical model for the temperature dependence of silicon refractive index from O to C band based on waveguide measurements," Opt. Express **27**, 27229–27241 (2019).
34. D.-X. Xu, A. Delâge, R. McKinnon, M. Vachon, R. Ma, J. Lapointe, A. Densmore, P. Cheben, S. Janz, and J. H. Schmid, "Archimedean spiral cavity ring resonators in silicon as ultra-compact optical comb filters," Optics express **18**, 1937–1945 (2010).
35. E. Vargas-Rodríguez and H. N. Rutt, "Analytical method to find the optimal parameters for gas detectors based on correlation spectroscopy using a Fabry–Perot interferometer," Appl. Opt. **46**, 4625–4632 (2007).
36. P. Cheben, R. Halir, J. H. Schmid, H. A. Atwater, and D. R. Smith, "Subwavelength integrated photonics," Nature **560**, 565–572 (2018).
37. P. Cheben, D.-X. Xu, S. Janz, and A. Densmore, "Subwavelength waveguide grating for mode conversion and light coupling in integrated optics," Opt. Express **14**, 4695–4702 (2006).
38. J. Witzens, "High-Speed Silicon Photonics Modulators," Proceedings of the IEEE **106**, 2158–2182 (2018).
39. N. Jovanovic, C. Schwab, O. Guyon, J. Lozi, N. Cvetojevic, F. Martinache, S. Leon-Saval, B. Norris, S. Gross, D. Doughty, T. Currie, and N. Takato, "Efficient injection from large telescopes into single-mode fibres: Enabling the era of ultra-precision astronomy," A&A **604**, A122 (2017).
40. T. Zhu, Y. Hu, P. Gatkine, S. Veilleux, J. Bland-Hawthorn, and M. Dagenais, "Arbitrary on-chip optical filter using complex waveguide Bragg gratings," Appl. Phys. Lett. **108**, 101104 (2016).
41. Y. Hu, S. Xie, J. Zhan, Y. Zhang, S. Veilleux, and M. Dagenais, "Integrated Arbitrary Filter With Spiral Gratings: Design and Characterization," Journal of Lightwave Technology **38**, 4454–4461 (2020).